\documentclass
[preprint,10pt,nofootinbib,showpacs,nobibnotes,superscriptaddress,balancelastpage,
,titlepage,twocolumn,pre]{revtex4}%
\usepackage{amsfonts}
\usepackage{amsmath}
\usepackage{amssymb}
\usepackage{graphicx}%
\providecommand{\U}[1]{\protect\rule{.1in}{.1in}}

\begin{document}
\preprint{ }
\title{Remarks about the thermostatistical description of the HMF model\\Part II: Phenomenology of Relaxation Dynamics}
\author{L. Velazquez}
\affiliation{Departamento de F\'{\i}sica, Universidad de Pinar del R\'{\i}o, Mart\'{\i}
270, Esq. 27 de Noviembre, Pinar del R\'{\i}o, Cuba.}
\author{F. Guzm\'{a}n}
\affiliation{Departamento de F\'{\i}sica Nuclear, Instituto Superior de Tecnolog\'{\i}a y
Ciencias Aplicadas, Carlos III y Luaces, Plaza, La Habana, Cuba.}
\date{\today}

\begin{abstract}
After a general overview of some features of the relaxation dynamics of the
Hamiltonian Mean Field model, its equilibrium thermodynamic properties are
used to rephrase the out-of-equilibrium regime for energies below the critical
point $u_{c}=0.75$ in terms of an effective dynamical coexistence between a
clustered and a gaseous phases, whose existence could be associated to the
large relaxation times observed when $u_{1}<u<u_{c}$, with $u_{1}=0.5$.
Starting from the hypothesis that the \textit{parametric resonance} is the
microscopic mechanism allowing the energetic interchange between the particles
during the collisional regime, a phenomenological Fokker-Planck equation based
on a Langevin equation with a multiplicative noise is proposed in order to
describe the collisional relaxation of this system towards its final
equilibrium, which supports the following dependence of the collisional
relaxation timescale $\tau_{cr}=\tau_{0}N\equiv\sqrt{IN/g}$.

\end{abstract}

\pacs{05.20.Gg; 05.20.-y}
\maketitle

\section{Introduction}

Despite of the Hamiltonian Mean Field (HMF) model
\cite{kk,pichon,inaga,ant,lat1,lat2,lat3,lat4,vrt,zanette,dauxois,yamaguchi,ybd,chava,chava2,chava3,bouchet1,bouchet2}
is a ferromagnetic toy model enough amenable for allowing an accurate
numerical and analytical characterization, it exhibits many features observed
in more realistic long-range interacting systems such as: violent relaxation,
persistence of metaequilibrium states, slow collisional relaxation, phase
transition, anomalous diffusion, etc. These features support the reason why it
can be considered as a paradigmatic toy model of the real long-range
interacting systems \cite{chava}.

The incidence of long-range interactions does not allow to divide this system
into independent subsystems even in the thermodynamic limit. This fact
evidences its intrinsic nonextensive nature, which distinguishes the HMF model
from other ferromagnetic models with short-range interactions despite they
share many analogies in most of the equilibrium thermodynamic properties
\cite{chava}. The nonextensivity of the HMF model is particularly important to
understand its nontrivial dynamical behavior, which is still an open problem
attracting much attention in the last years
\cite{dauxois,yamaguchi,ybd,chava,chava2,chava3,bouchet1,bouchet2}.

The present paper of this series \cite{vel.HMF.1} is the first work devoted to
study the dynamical behavior of the HMF model with a special emphasis on the
relaxation processes undergone by this system during the evolution towards the
thermodynamic equilibrium. We begin performing a general characterization of
the most important features of the microscopic dynamics by taking into
consideration the results obtained from the analysis of its equilibrium
thermodynamics \cite{vel.HMF.1}. Such a study possibilities us to conjecture
the equilibration mechanism leading this system to its final relaxation, which
is here used to propose a phenomenological approach of the collisional evolution.

\section{General characterization\label{dynamics}}

\subsection{Numerical computation of microscopic dynamics and the Vlasov
equation}

According to the Hamiltonian:%
\begin{equation}
H_{N}=\sum_{i=1}^{N}\frac{1}{2I}L_{i}^{2}+\frac{1}{2}g\sum_{i=1}^{N}\sum
_{j=1}^{N}\left[  1-\cos\left(  \theta_{i}-\theta_{j}\right)  \right]  ,
\label{H}%
\end{equation}
the motion equations of the HMF model are given by:%

\begin{equation}
\dot{\theta}_{i}=\frac{1}{I}L_{i},~\dot{L}_{i}=gN\left(  m_{y}\cos\theta
_{i}-m_{x}\sin\theta_{i}\right)  \text{,} \label{motion equations}%
\end{equation}
for $i=1,2,...N$, where $m_{x}$ and $m_{y}$ are the Cartesian components of
the magnetization vector $\mathbf{m=}\left(  m_{x},m_{y}\right)  =\left(
\sum_{i=1}^{N}\mathbf{m}_{i}\right)  /N\mathbf{\ }$where $\mathbf{m}%
_{i}=\left(  \cos\theta_{i},\sin\theta_{i}\right)  $.

From a dynamical viewpoint, the HMF model represents the long-range
interacting version of harmonic oscillators system \cite{chava} since the
individual dynamics of a given rotator can be easily rephrased as the
dynamics\ of the \textit{mathematical pendulum}:%

\begin{equation}
\ddot{\theta}_{i}+m\sin\left(  \theta_{i}-\theta\right)  =0. \label{oscilator}%
\end{equation}
We have used the polar representation of the magnetization vector
$\mathbf{m}=m\left(  \cos\theta,\sin\theta\right)  $ and the characteristic
units for time and momentum:%

\begin{equation}
\tau_{0}=\sqrt{\frac{I}{gN}}\text{ and }L_{0}=\sqrt{IgN},
\label{characteristics units}%
\end{equation}
in terms of the moment of inertia $I$ and the coupling constant $g$.
Obviously, the characteristic time unit $\tau_{0}$ provides the timescale for
the evolution of each rotator, and hence, $\tau_{0}$ is the characteristic
microscopic timescale. The use of characteristic unit $L_{0}$ allows to
express the total energy per particle $u=k+\frac{1}{2}\left(  1-\mathbf{m}%
^{2}\right)  $ and the kinetic energy per particle $k=K/N=\left(  \sum
_{i=1}^{N}\frac{1}{2}p_{i}^{2}\right)  /N$, with $p_{i}=L_{i}/L_{0}$, in terms
of the characteristic unit $\varepsilon_{0}=E_{0}/N=gN$ introduced in our
previous paper \cite{vel.HMF.1}.

Another convenient way to rewrite the dynamics of the HMF model
(\ref{motion equations}) follows from the introduction of the tridimensional
vectors $\mathbf{m}_{i}=\left(  \cos\theta_{i},\sin\theta_{i},0\right)  $ and
$\mathbf{o}_{k}=\left(  0,0,p_{i}\right)  $:%

\begin{equation}
\mathbf{\dot{m}}_{i}=\mathbf{o}_{i}\times\mathbf{m}_{i},~\mathbf{\dot{o}}%
_{i}=\mathbf{m}_{i}\times\mathbf{m}\text{.} \label{vectorial form}%
\end{equation}
where the magnetization vector $\mathbf{m}$ obeys the following dynamics:%

\begin{equation}
\mathbf{\dot{m}}=\mathbf{\Upsilon}=\frac{1}{N}\sum_{i=1}^{N}\mathbf{o}%
_{i}\times\mathbf{m}_{i}, \label{magnetization dynamics}%
\end{equation}
being $\mathbf{\Upsilon}$ the magnetization rate of change.

A simple inspection of Eqs.(\ref{oscilator}), (\ref{vectorial form}) and
(\ref{magnetization dynamics}) allows to understand that the dynamical
phenomenology the HMF model can be explained from the conjugation of the
dynamical features of the mathematical pendulum and the character of the
magnetization vector $\mathbf{m}$ evolution. Since the individual terms
$\mathbf{o}_{k}\times\mathbf{m}_{k}$ in the\ magnetization rate of change
$\mathbf{\Upsilon}$ have a undefined signature, the dynamical evolution of
magnetization vector crucially depends on the correlations among the
individual rotators evolution. If present, the magnetization experiences large
variations in a timescale comparable with the microscopic time $\tau_{0}$,
where takes place a very effective energy interchange among the rotators. This
dynamical regime characterized by the existence of a collective motion
completely analogue to the one operating in the astrophysical systems, which
is usually referred as \textit{violent relaxation} \cite{bin} and very-well
described by the \textit{Vlasov dynamics}:%
\begin{equation}
\left(  \frac{\partial}{\partial t}+p\frac{\partial}{\partial\theta
}+\mathbf{m}\left(  \theta\right)  \times\mathbf{m}\left[  f\right]
\frac{\partial}{\partial p}\right)  f\left(  \theta,p;t\right)  =0, \label{vd}%
\end{equation}
with $\mathbf{m}\left[  f\right]  =\int\mathbf{m}\left(  \theta\right)
f\left(  \theta,p;t\right)  d\theta dp$, which accounts the
\textit{collisionless} regime of the one-body distribution function $f\left(
\theta,p;t\right)  $. This fast relaxation regime finishes after arriving at
certain stable \textit{quasi-stationary state }(QSS)\textit{\ }of
Eq.(\ref{vd}), \textit{ }$f\left(  \theta,p;t\right)  \rightarrow
f_{QSS}\left(  \theta,p\right)  =F\left[  \varepsilon\left(  p,\theta\right)
\right]  $, where $\varepsilon\left(  p,\theta\right)  =\frac{1}{2}%
p^{2}-\mathbf{m\cdot m}\left(  \theta\right)  $ is the energy of an individual
rotator. Hereafter, the dynamical evolution depends on \textit{collisional
effects that }decrease with the system size $N$, and hence, the system evolves
throughout stable QSSs towards its final equilibration
\cite{dauxois,yamaguchi,ybd,chava,chava2,chava3,bouchet1,bouchet2}.

During this quasi-stationary regime, the individual rotators becomes almost
non correlated, since the two-body correlation function probably survives at
the $1/N$ approximation as its microcanonical estimate, Eq.(58) of
ref.\cite{vel.HMF.1}. Therefore, their short-time dynamics can be described as
free motions of a mathematical pendulum. The vector $\mathbf{\Upsilon}$
experiences zero mean fluctuations in the microscopic timescale $\tau_{0}$
whose amplitudes decrease with the system size as $1/\sqrt{N}$, provoking in
this way a very small fluctuating behavior of the magnetization vector
$\mathbf{m}$ around certain mean value. Such fluctuations constitute the
microscopic relaxation mechanism allowing the energetic interchange among the
rotators, which reduces its effectiveness with the increasing of the system
size $N$. This last observation clarifies the reason why the characteristic
relaxation timescale of such a collisional regime $\tau_{cr}\sim\tau_{eq}$
grows with $N$.

The above picture is easily verified by means of microcanonical numerical
computation of dynamics (\ref{motion equations}). For comparison purposes, we
also carry out the numerical integration of the collisionless dynamics
(\ref{vd}) by imposing a truncation of distribution function $f\left(
\theta,p;t\right)  \equiv0$ for $\left\vert p\right\vert \geq p_{c}$ and
introducing a second-order finite differences scheme as follows:%
\begin{equation}
\frac{\partial f_{ij}}{\partial t}+p_{j}\Delta_{\theta}f_{ij}+\mathbf{m}%
_{i}\times\mathbf{m}\left[  f\right]  \Delta_{p}f_{ij}=0, \label{fd-scheme}%
\end{equation}
where $f_{i,j}=f\left(  \theta_{i},p_{j}\right)  $ with $\theta_{i}=ih_{1}$
and $p_{j}=jh_{2}$, being $i=\left[  1,2,\ldots,N_{1}+1\right]  $, $j=\left[
-N_{2}-1,\ldots,N_{2}+1\right]  $, $h_{1}=2\pi/\left(  N_{1}+1\right)
\,$\ and $h_{2}=p_{c}/\left(  N_{2}+1\right)  $, with the boundary
conditions:
\begin{equation}
f_{N_{1}+1,j}=f_{1,j},~~f_{i,\pm\left(  N_{2}+1\right)  }=0.
\end{equation}
Besides, magnetization is rephrased as $\mathbf{m}\left[  f\right]
=h_{1}h_{2}\sum_{ij}\mathbf{m}_{i}f_{ij}$ with $\mathbf{m}_{i}=\mathbf{m}%
\left(  \theta_{i}\right)  $, and the partial numerical derivatives are given
by $\Delta_{\theta}f_{ij}=\left(  f_{i+1,j}-f_{i,j}\right)  /h_{1}$ and
$\Delta_{p}f_{ij}=\left(  f_{i,j+1}-f_{i,j-1}\right)  /2h_{2}$. Numerical
integrations of the microscopic dynamics (\ref{motion equations}) and the
scheme (\ref{fd-scheme}) with $N_{1}=101$ and $N_{2}=200$ where performed by
using fourth-order Runge-Kutta method with constant timesteps $\delta
t_{m}=0.05$ and $\delta t_{v}=0.02$ respectively, which ensure a good enough
conservation of the energy per particle $u$. All microscopic simulations
starts from \textit{water bag initial conditions} (WB): $\left\vert
\mathbf{m}\left(  t=0\right)  \right\vert =1$ $\left(  \forall i,~\theta
_{i}=0\right)  $ with a random uniform distribution for rotators momenta
$\left[  -p_{\ast},p_{\ast}\right]  $\thinspace\ with $p_{\ast}=\sqrt{6u}$, an
unstable initial condition which can be considered as microscopic
configuration highly ordered. In order to avoid abrupt changes in the partial
numerical derivatives $\Delta_{\theta}f_{ij}$ and $\Delta_{p}f_{ij}$, the WB
initial conditions of the microscopic dynamics were approximated by the
following initial condition for the one-body distribution function:
\[
f_{wb}\left(  \theta,p\right)  \propto\frac{1}{1+\exp\left[  \lambda_{\theta
}\left(  \left\vert \theta\right\vert -\theta_{\ast}\right)  \right]  }%
\frac{1}{1+\exp\left[  \lambda_{p}\left(  \left\vert p\right\vert -p_{\ast
}\right)  \right]  },
\]
with $\lambda_{\theta}=\lambda_{p}=50$ and momentum cutoff $p_{c}=1.5p_{\ast}%
$. The small parameter $\theta_{\ast}$ used here ensures a precision of the
initial magnetization as $\delta\left\vert \mathbf{m}\left(  0\right)
\right\vert \sim10^{-3}$. As in many studies
\cite{lat3,lat4,zanette,dauxois,yamaguchi,ybd}, the present calculations were
performed at $u=0.69$, an energy value corresponding at the equilibrium to the
ferromagnetic state close to the critical point $u_{c}=0.75$ of the continuous
phase transition.%

\begin{figure}
[t]
\begin{center}
\includegraphics[
height=2.6082in,
width=3.5405in
]%
{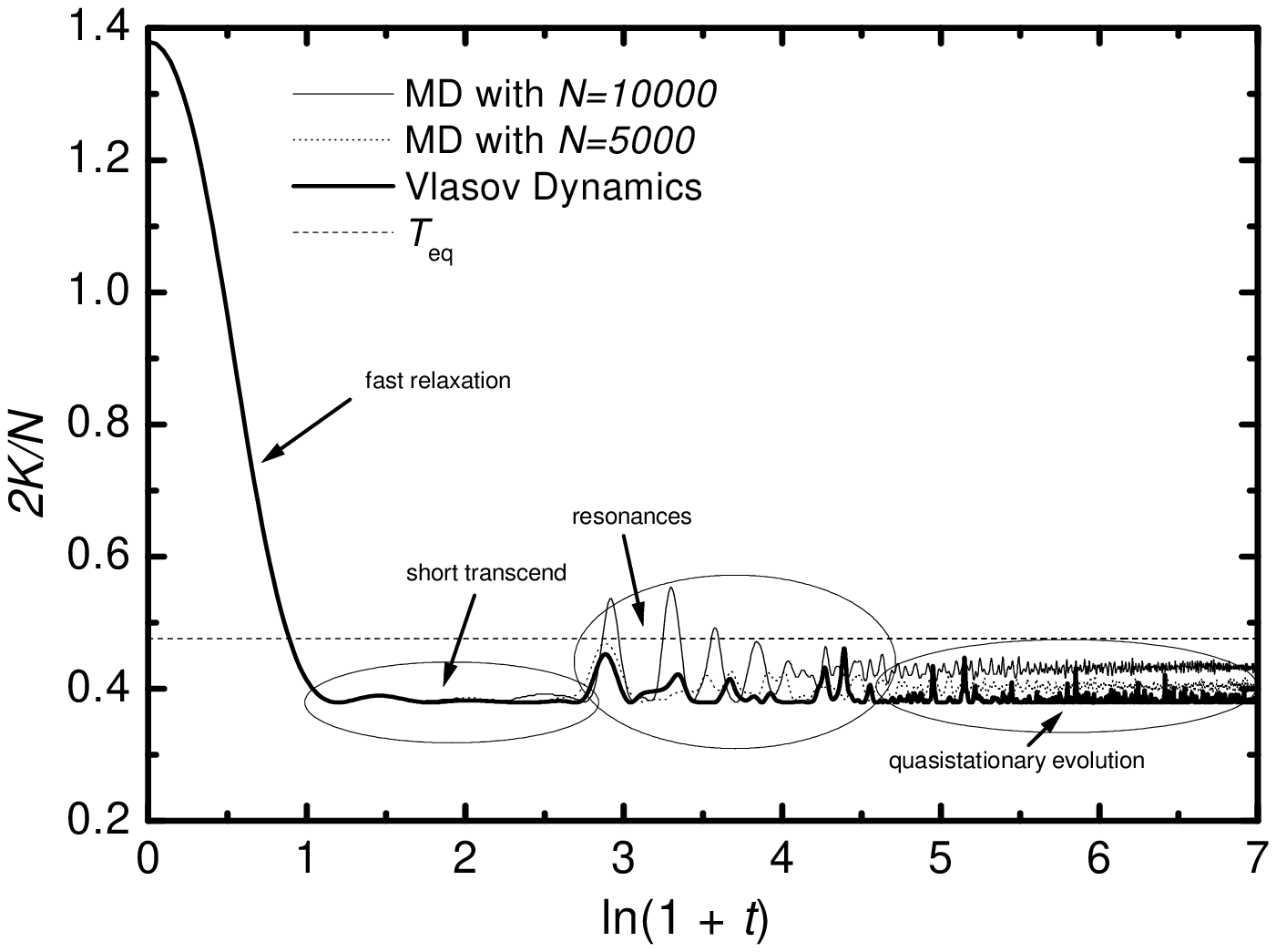}%
\caption{Evolution of the nonequilibrium temperature $T_{D}$ obtained from the
microscopic dynamics (MD) with $N=5000$ and $10000$, which can be compared to
the Vlasov collisionless dynamics. Horizontal scale represents the function
$\ln\left(  1+t\right)  $ in order to exhibit the whole simulation time
interval by using a linear scale for small $t$ and in logarithmic scale when
$t\gg1$.}%
\label{micdyn.eps}%
\end{center}
\end{figure}

FIG.\ref{micdyn.eps} shows the evolution of the \textit{nonequilibrium
temperature} $T_{D}=2K/N$ (two times the kinetic energy per particle) obtained
from two individual trajectories of (\ref{motion equations}) with $u=0.69$ for
$N=5000$ and $10000$ respectively. Notice that the temporal dependences of
$T_{D}$ derived from the microscopic dynamics can be compared to the evolution
of this same quantity obtained from the numerical implementation of the Vlasov dynamics.

Size effects are hardly appreciated at the first stages of the microscopic
dynamics where takes place a very fast relaxation followed by a short
transcend regime (the plateau in $T_{D}$), which finishing with another
relaxation regime where take place the development of resonances (peaks) with
the incidence of finite size effects. The fact that the immediate response\ of
the system can be accounted for by the Vlasov dynamics demonstrates the
existence of a collisionless evolution during this initial relaxation regime,
whose relaxation time $\tau_{vr}$ is comparable to the microscopic timescale
$\tau_{0}$, $\tau_{vr}\sim\tau_{0}$. FIG.\ref{micdyn.eps} also shows that the
size effects and fluctuations are very important in the subsequent evolution
after the violent relaxation. The large transcend regime evidences the
occurrence of quasi-stationary evolution where the nonequilibrium temperature
$T_{D}$ remains a long time fluctuating around a nonequilibrium value (compare
it with the equilibrium temperature $T_{eq}=0.476$ represented as a horizontal line).

\begin{figure}
[t]
\begin{center}
\includegraphics[
height=2.5382in,
width=3.5111in
]%
{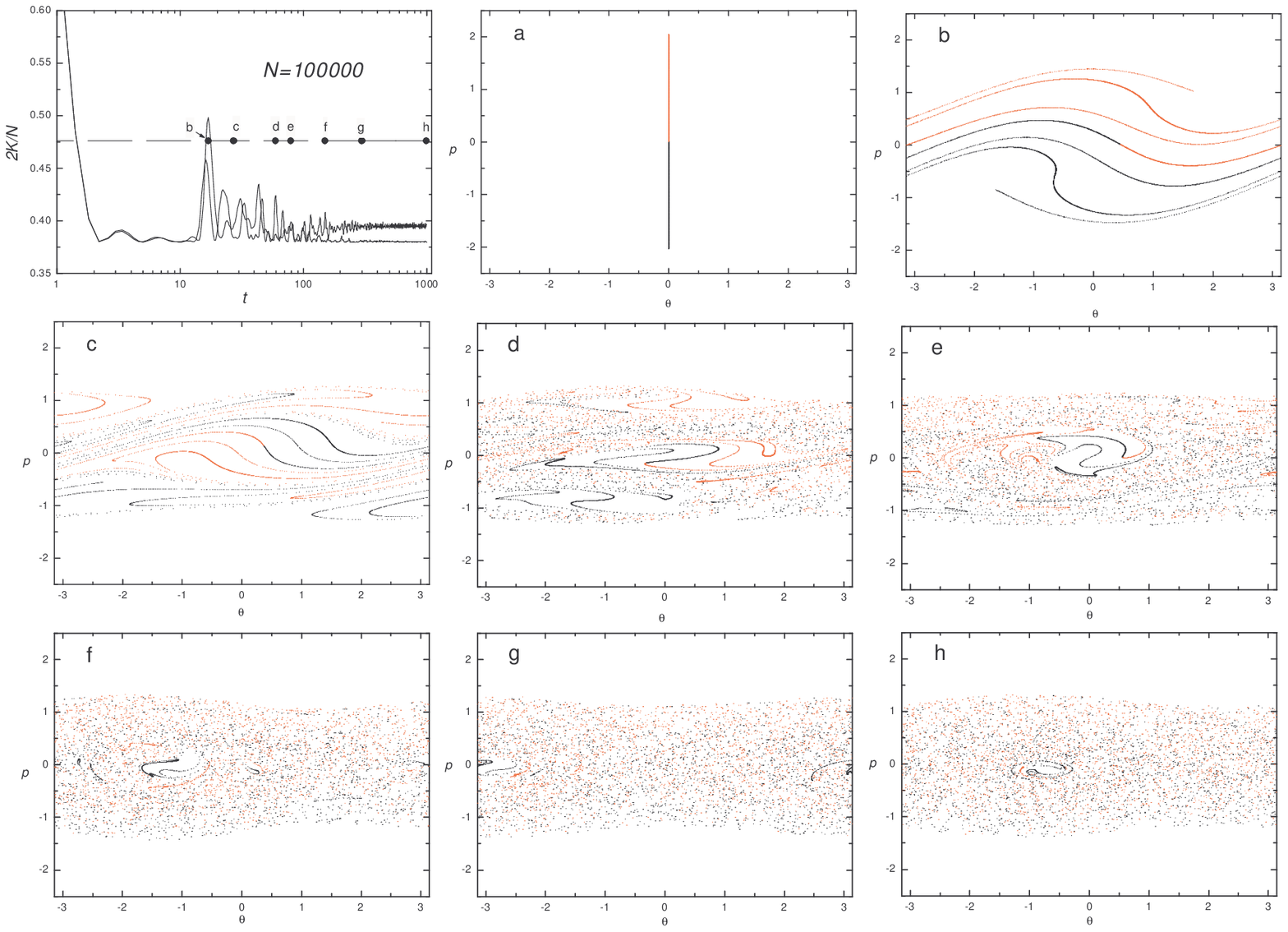}%
\caption{First panel: Dynamical evolution of the nonequilibrium temperature
associated with two realizations of the wb initial conditions with $N=100000$,
where it could be appreciated a dynamical sensibility to the initial
conditions. Panels (a-h) display different instants (indicated in the first
panel) of the dynamical evolution of the system in the $\mu$-space, where it
is revealed the formation and destruction of filamentary structures. Black and
red points are use to illustrate the mixing of microscopic dynamics.}%
\label{qpspace2.eps}%
\end{center}
\end{figure}

\begin{figure}
[t]
\begin{center}
\includegraphics[
height=2.3177in,
width=3.5111in
]%
{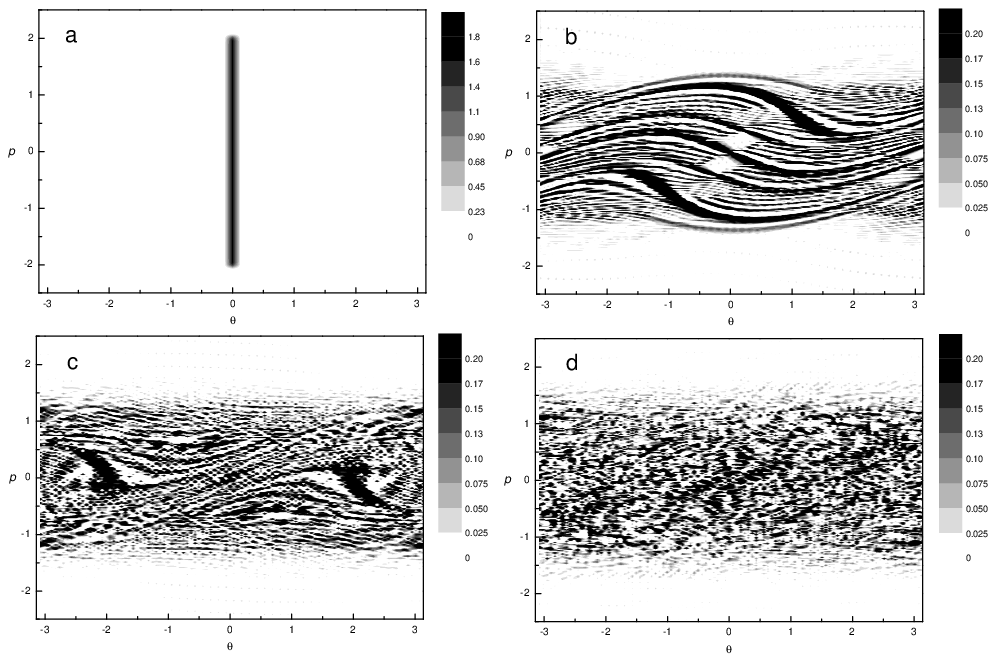}%
\caption{Collisionless evolution of one-body distribution function obtained
from the numerical integration of the finite differences scheme
(\ref{fd-scheme}) where it can be also appreciated the formation and
destruction of filamentary structures. We illustrate here the first four
instants shown in the FIG.\ref{qpspace2.eps}.}%
\label{vp1.eps}%
\end{center}
\end{figure}

The evolution of a given realization of the microscopic dynamics in the $\mu
$-space $\left(  \theta,p\right)  $ is shown in FIG.\ref{qpspace2.eps} for
$N=100000$. It illustrates how the highly ordered initial state is
progressively destroyed by the incidence of chaoticity and mixing of the
collisionless regime. Such mechanisms manifest in the development of
filamentary structures, which disappear at a coarsed grained scale leading
thus to the establishment of the stable QSS. This process is also displayed in
FIG.\ref{vp1.eps}, but this time, the evolution of the one-body distribution
function obtained from the finite differences scheme (\ref{fd-scheme}) where
it could be also noticed the formation and destruction of filamentary structures.%

%

Such a large transcend regime could be identify with the presence of a stable
QSS of the Vlasov dynamics (\ref{vd}), which actually evolves in the
collisional timescale $\tau_{cr}\gg\tau_{0}$. This fact has been shown
throughout extensive calculation of the average of the nonequilibrium
temperature $T_{D}$ with $N=10^{2}-10^{4}$ rotators over $\sim10^{3}-10^{4}$
realizations of microscopic dynamics, where it is shown a growing of the
relaxation timescale $\tau_{eq}$ with $N$ \cite{dauxois,yamaguchi,ybd}. As
already pointed out theoretically and observed in many numerical studies, the
nature and the relaxation time of such a collisional quasi-stationary
evolution depend on the initial conditions.

\begin{figure}
[t]
\begin{center}
\includegraphics[
height=2.6204in,
width=3.5405in
]%
{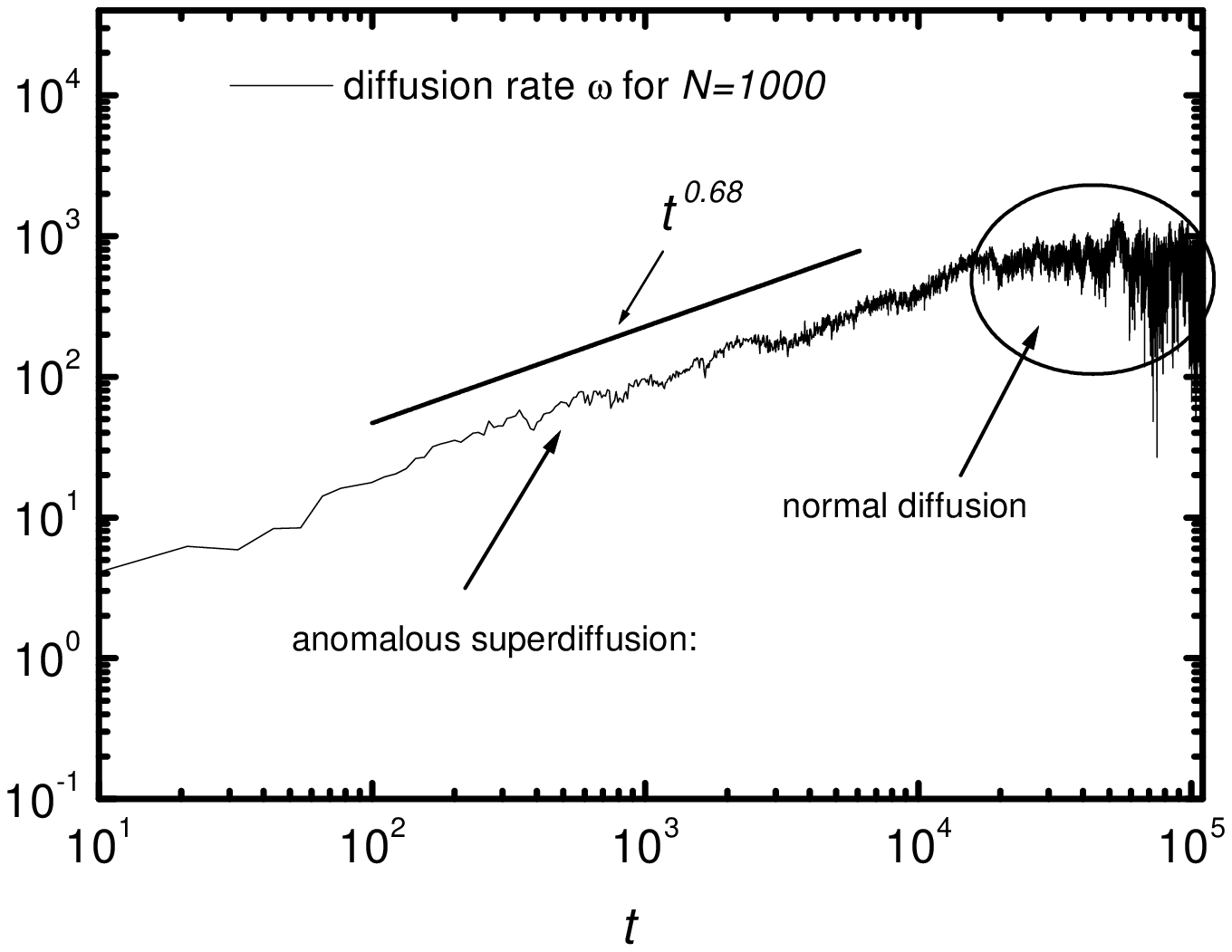}%
\caption{Evolution of diffusion rate $\omega\left(  t\right)  $ obtained from
the numerical simulation of the microscopic dynamics with $N=1000$, which
evidences the existence of a crossover time at $\tau_{cross}\sim10^{4}$ where
the anomalous superdiffusion with effective exponent $\alpha\simeq1.68$
turns\ a normal diffussion.}%
\label{anormal.diffusion.eps}%
\end{center}
\end{figure}

The qualitative features of system evolution are also captured by the
\textit{diffusion rate} $\omega\left(  t\right)  =d\Delta_{\theta}\left(
t\right)  /dt$ defined from the square dispersion $\Delta_{\theta}\left(
t\right)  $:
\begin{equation}
\Delta_{\theta}\left(  t\right)  =\frac{1}{2N}\sum_{i=1}^{N}\left[  \theta
_{i}\left(  t\right)  -\theta_{i}\left(  0\right)  \right]  ^{2}%
,\label{diffusion rate}%
\end{equation}
which provides a measure of the average correlation function:
\begin{equation}
\omega\left(  t\right)  =\frac{1}{N}\sum_{i=1}^{N}\omega_{i}\left(  t\right)
,\text{with }\omega_{i}\left(  t\right)  \equiv\int_{0}^{t}L_{i}\left(
t\right)  L_{i}\left(  t-\tau\right)  d\tau,
\end{equation}
and accounts for the character of the diffusion regimes undergone by the
system\footnote{A regime with $\Delta_{\theta}\left(  t\right)  \propto
t^{\alpha}$ exhibits a normal diffusion when $\alpha=1$, ballistic diffusion
$\alpha=2$, and \textit{anomalous diffusion} when $1<\alpha<2$
(superdiffusion) or $0<\alpha<1$ (sub-diffusion).}. Dynamical studies revealed
that after a brief ballistic regime with $\alpha=2$, the system exhibits an
\textit{anomalous superdiffusion} during the quasi-stationary evolution. The
long time calculation of the diffusion rate illustrated in
FIG.\ref{anormal.diffusion.eps} allows to estimate the exponential constant
$\alpha\simeq1.68$ for the leading behavior of the square dispersion with
$N=1000$. Notice the existence of a crossover time $\tau_{cross}\sim10^{4}$
where the diffusion regime turns \textit{normal}, which can be related to the
final evolution of the collisional regime towards the achievement the of the
Boltzmann-Gibbs equilibrium. Despite the incidence of size effects in the
crossover time $\tau_{cross}$ and the exponent $\alpha$, the general behavior
of the diffusional regime described above is consistent for different system
sizes $N$.%

\subsection{Relaxation of two--body correlation function}

By denoting $\delta\left[  x-x_{i}\left(  t\right)  \right]  \equiv
\delta\left[  \theta-\theta_{i}\left(  t\right)  \right]  \delta\left[
p-p_{i}\left(  t\right)  \right]  $ with $x=\left(  \theta,p\right)  $ and the
average over many realizations of the microscopic dynamics $\left\langle
\cdot\right\rangle $, we are able to introduce the one-body and the two-body
distribution functions:
\begin{align}
&
\begin{array}
[c]{c}%
f\left(  x;t\right)  =\left\langle \frac{1}{N}\sum_{i=1}^{N}\delta\left[
x-x_{i}\left(  t\right)  \right]  \right\rangle ,~~
\end{array}
\label{2-dff}\\
&
\begin{array}
[c]{c}%
f^{\left(  2\right)  }\left(  x_{1},x_{2};t\right)  =\frac{2}{N\left(
N-1\right)  }\left\langle \sum_{i>j}^{N}\delta\left[  x_{1}-x_{i}\left(
t\right)  \right]  \delta\left[  x_{2}-x_{j}\left(  t\right)  \right]
\right\rangle ,
\end{array}
\nonumber
\end{align}
as well as the two-body correlation function $g\left(  x_{1},x_{2};t\right)  $
as follows:%
\begin{equation}
f^{\left(  2\right)  }\left(  x_{1},x_{2};t\right)  =f\left(  x_{1};t\right)
f\left(  x_{2};t\right)  +g\left(  x_{1},x_{2};t\right)  . \label{2-cf}%
\end{equation}
These distributions functions can be used to define the average $\left\langle
a\right\rangle $ and the correlation function $c_{a}$ of a microscopic
observable $a\left(  x\right)  =a\left(  \theta,p\right)  $ of an individual
rotator:
\begin{align}
\left\langle a\right\rangle  &  \equiv\int a\left(  x\right)  f\left(
x;t\right)  dx,\label{1-mean}\\
c_{a}  &  =\int a\left(  x_{1}\right)  a\left(  x_{2}\right)  g\left(
x_{1},x_{2};t\right)  dx_{1}dx_{2}, \label{2-corr}%
\end{align}
with $dx=d\theta dp$, which represent two indicators of \ their corresponding
dynamical evolutions.%

\begin{figure}
[t]
\begin{center}
\includegraphics[
height=4.0062in,
width=3.2128in
]%
{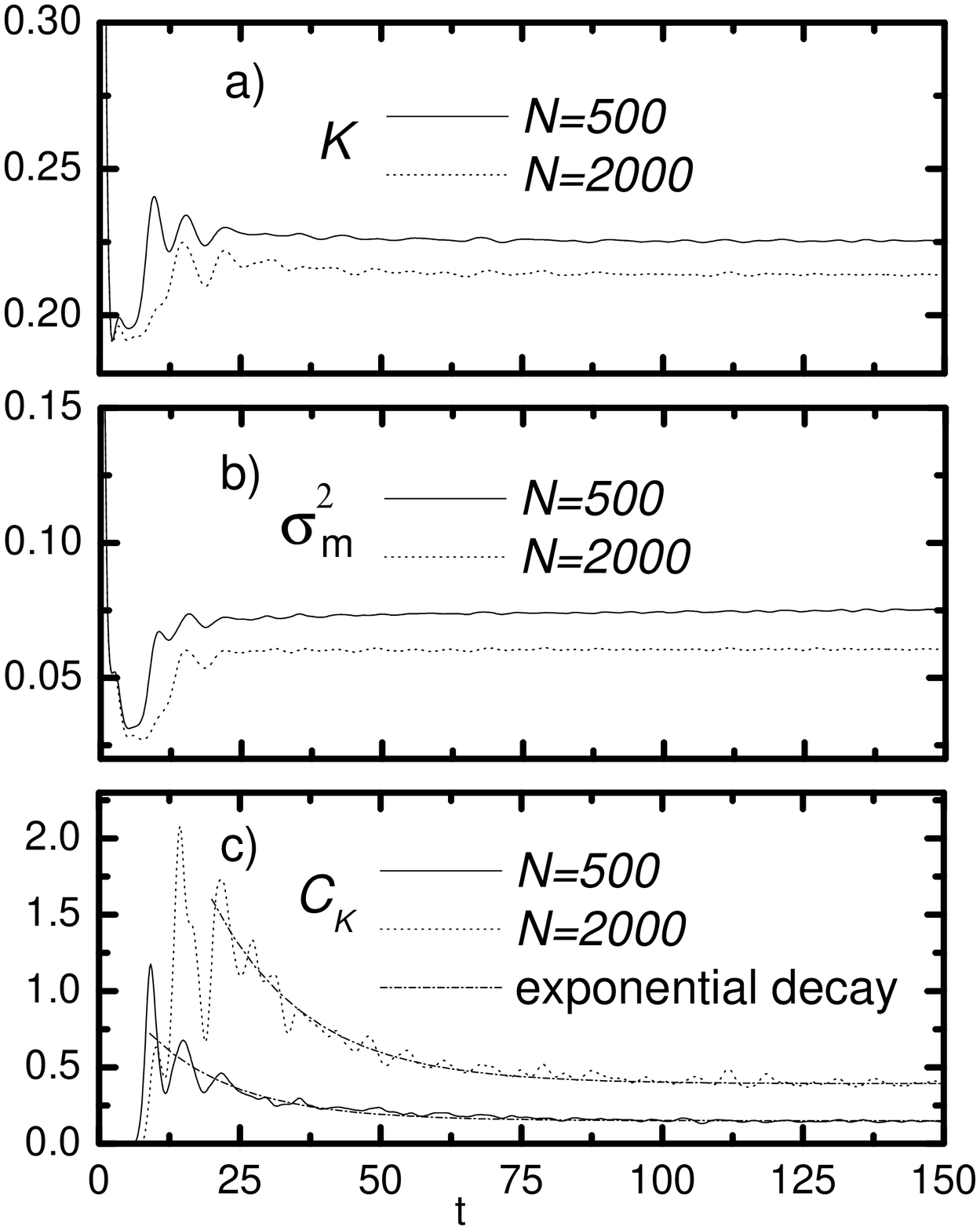}%
\caption{Short-time evolutions of (a) the kinetic energy $\left\langle
K\right\rangle $, (b) the square dispersion of individual rotators $\sigma
_{m}^{2}$ and (c) the correlation function $C_{K}$ obtained from a statistical
average over 1500 trajectories with $N=500$ and $2000$. The exponential decay
of the correlation function $C_{K}$ evidences that the two-body correlation
function $g\left(  x_{1},x_{2};t\right)  $ undergoes \textit{a relaxation
process} \textit{in the same timescale of the violent relaxation} of the
one-body distribution function $f\left(  x;t\right)  $.}%
\label{damping.eps}%
\end{center}
\end{figure}

The correlation function of the kinetic energy $c_{K}$ can be derived from the
square dispersion of the kinetic energy per particle $\sigma_{K}%
^{2}=\left\langle K^{2}\right\rangle -\left\langle K\right\rangle ^{2}$
obtained after several realizations of the microscopic dynamics as follows:
\begin{equation}
c_{K}=\frac{1}{N-1}\left\{  N\sigma_{K}^{2}-\sigma_{m}^{2}\right\}
\equiv\frac{1}{N-1}C_{K}.
\end{equation}
where $\sigma_{m}^{2}=\frac{1}{4}\left\{  \left\langle p^{4}\right\rangle
-\left\langle p^{2}\right\rangle ^{2}\right\}  $ is the square dispersion of
the kinetic energy of an individual rotator, which in an analogous way as
$\left\langle K\right\rangle $ only depends on the one-body distribution
function. The short-time evolutions of the functions $\left\langle
K\right\rangle $, $\sigma_{m}^{2}$ and $C_{K}$ obtained from the average of
microscopic dynamics with $N=500$ and $2000$ over 1500 trajectories are
displayed in FIG.\ref{damping.eps}.

While $\left\langle K\right\rangle $ and $\sigma_{m}^{2}$ account for the
violent relaxation undergone by the one-body distribution function $f\left(
\theta,p;t\right)  $ during the first stages of the system evolution, the
correlation function $C_{K}$ evidences also the \textit{occurrence of\ a
relaxation process for the two-body correlation function} $g\left(  \theta
_{1},p_{1},\theta_{2},p_{2};t\right)  $ \textit{in the same timescale}, where
any two uncorrelated particles at the initial state become correlated after
the incidence of such a relaxation process. Despite the existence of finite
size effects, numerical simulations with different $N$ confirm the consistence
of this observation. We shall show in a forthcoming paper that this result can
be easily justified in terms of the well-known \textit{BBGKY hierarchy}.
Although simple, this latter behavior has a remarkable importance in
understanding the\ collisional relaxation of this model system and the
development of appropriate kinetic equations for the one-body distribution
function $f\left(  \theta,p;t\right)  $.

\subsection{Dynamical phase coexistence}

It is well-known that the existence of the long-range order leads to a very
slow relaxation in the neighborhood of the critical point. However, most of
works devoted to the numerical study of dynamical aspects of the HMF model
also revealed the existence of very large relaxation times during the
superdifussional regime within the energetic region $u_{1}=0.5<u\leq
u_{c}=0.75$ \cite{ant,lat1,lat2,lat3,lat4,vrt,zanette,yamaguchi,ybd}. It is
remarkable that the above energetic region is precisely the same one where the
microcanonical susceptibility $\chi_{m}\left(  u;b=0\right)  $ experiences a
large increasing below of the critical point $u_{c}$ (see in FIG.2 of the
previous paper \cite{vel.HMF.1}). This observation suggests that the
microcanonical results and these anomalous dynamical behaviors could be
closely related.

While the thermodynamical characterization of this model shows the features of
the typical second-order phase transitions, from the viewpoint of its
out-of-equilibrium dynamics, the HMF model also shows certain analogy with the
\textit{phase coexistence phenomenon}. The present interpretation is naturally
arisen from characterization of the rotator individual dynamics as a
mathematical pendulum motion: when the pendulum energy $\varepsilon\left(
\theta,p\right)  =\frac{1}{2}p^{2}-m\cos\theta<m$, the particle exhibits an
oscillatory regime around the equilibrium configuration, while for
$\varepsilon\left(  \theta,p\right)  >m$ the particle performs rotations in a
given direction. The first kind of motion characterizes the particles trapped
in the cluster existing when $u<u_{c}$, while the second one characterizes
those particles that are not bonded to the cluster, which is the typical
motion predominating in the homogeneous phase with $u>u_{c}$. \ We can
rephrase the microscopic dynamics by considering that all those rotators
satisfying the condition $\varepsilon\left(  \theta,p\right)  <m$ belong to a
\textit{dynamical clustered phase}, while the other with $\varepsilon\left(
\theta,p\right)  >m$ belong to a \textit{dynamical gaseous phase}. Such a
reinterpretation suggests the occurrence of a \textit{dynamical phase
coexistence} within the energetic region $u_{1}<u<u_{c}$.

The relative population $p$ of the clustered phase can be obtained from the
distribution function $f\left(  \theta,p;t\right)  $ as follows:%

\begin{equation}
1-p\left(  t\right)  =\int_{m}^{\infty}d\varepsilon F\left(  \varepsilon
;m,t\right)  , \label{def.p}%
\end{equation}
where $F\left(  \varepsilon;m,t\right)  $ is the energy distribution function:%
\begin{equation}
F\left(  \varepsilon;m,t\right)  =\int_{0}^{2\pi}d\theta\int_{-\infty
}^{+\infty}dp\delta\left\{  \varepsilon-\varepsilon\left(  \theta,p\right)
\right\}  f\left(  \theta,p;t\right)  .
\end{equation}
Since during collisional regime the distribution function exhibits a
quasi-stationary evolution throughout stable QSSs of Vlasov equation
(\ref{vd}), the distribution function can be taken as $f_{QSS}\left(
\theta,p;t\right)  \simeq f\left[  \varepsilon\left(  \theta,p\right)
;m,t\right]  $, which leads to the following form of the quasi-stationary
energy distribution function:%
\begin{equation}
F\left(  \varepsilon;m,t\right)  =T\left(  \varepsilon,m\right)  f\left(
\varepsilon;m,t\right)  , \label{qedf}%
\end{equation}
being $T\left(  \varepsilon,m\right)  $ the time period of the mathematical
pendulum with energy $\varepsilon$:%
\begin{equation}
T\left(  \varepsilon,m\right)  =4\int_{0}^{\theta_{m}}\frac{d\theta}{p\left(
\theta;\varepsilon,m\right)  }=4\int_{0}^{\theta_{m}}\frac{d\theta}%
{\sqrt{2\left(  \varepsilon+m\cos\theta\right)  }}, \label{T}%
\end{equation}
where $\theta_{m}$ is the positive turning point where $p\left(  \theta
_{m};\varepsilon,m\right)  =0$\ when $\varepsilon<m$, or $\theta_{m}\equiv\pi$
when $\varepsilon>m$.

The presence of the time period function $T\left(  \varepsilon,m\right)  $ in
Eq.(\ref{qedf}) demonstrates the existence of a \textit{pole} in
quasi-stationary energy distribution function at $\varepsilon=m$
($\lim_{\varepsilon\rightarrow m}T\left(  \varepsilon,m\right)  =\infty$),
which separates the energetic range of the system from the dynamical viewpoint
in a clustered phase when $\varepsilon<m$ and a gaseous phase for
$\varepsilon>m$. The incidence of such a lost of analyticity on the
quasi-stationary regime is much significant for energies $u\leq u_{c}$ close
to the critical point. The relatively large collisional relaxation times and
the anomalous superdiffusional observed in the interval $u_{1}<u<u_{c}$ should
be related in some way to this dynamical phase coexistence. Unfortunately, a
complete study of these question demands the knowledge of a kinetic equation
describing the collisional evolution of the system.%

\begin{figure}
[t]
\begin{center}
\includegraphics[
height=2.8106in,
width=3.513in
]%
{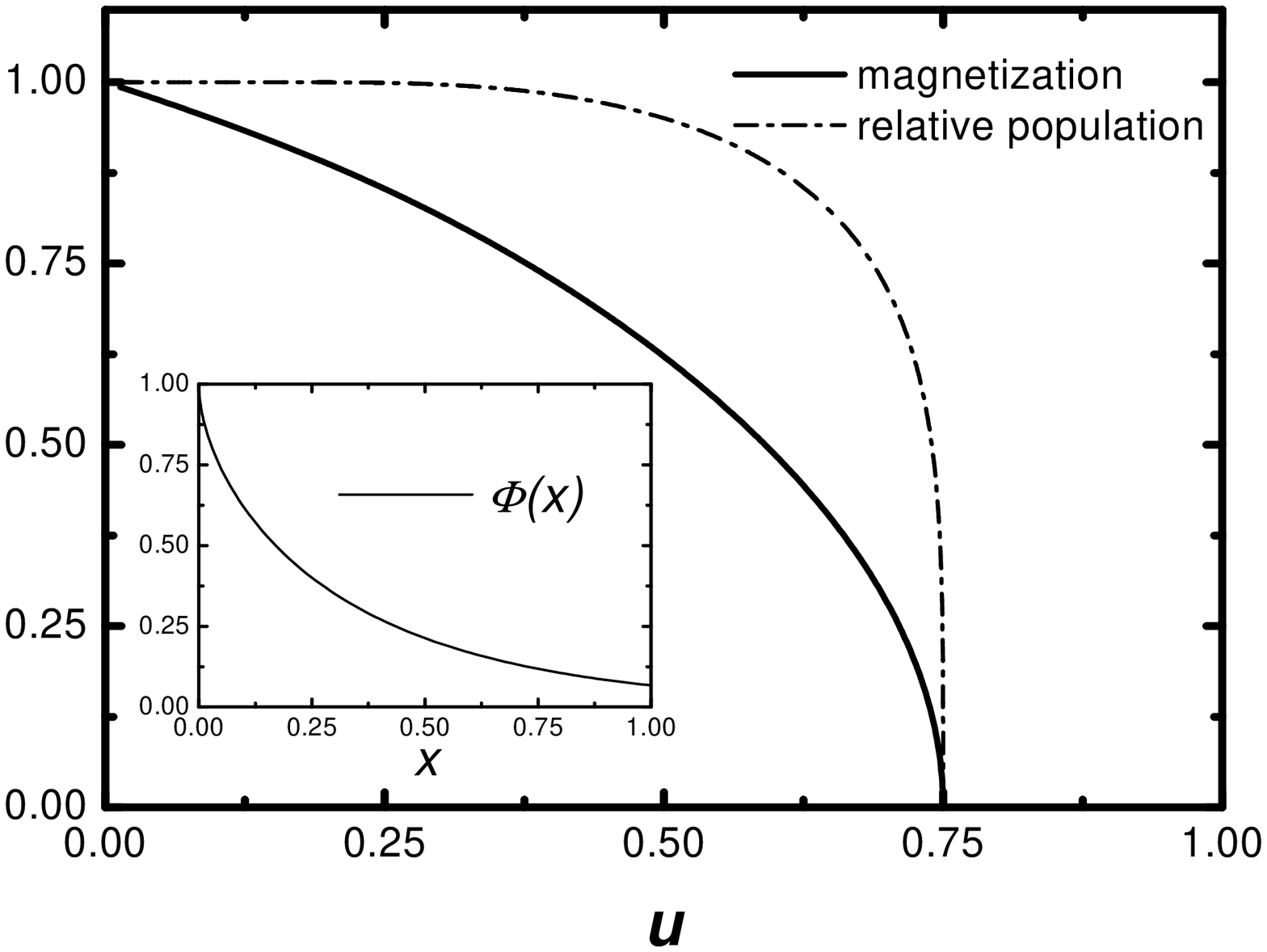}%
\caption{Energy dependence of the relative population $p$ of the clustered
phase and the magnetization $m$ (added for comparative purposes). Inserted
graph: the $x$ dependence of the function $\Phi\left(  x\right)  $. This
figure shows that the relative population of the clustered phase exhibits its
most significant variation within the energy interval $u_{1}<u\leq u_{c}$.}%
\label{population.eps}%
\end{center}
\end{figure}

Nevertheless, it could be useful to calculate the relative population $p$ of
the clustered phase during the thermodynamic equilibrium of the HMF model. The
integration of (\ref{def.p}) by using\ the Boltzmann-Gibbs distribution
function $f_{BG}\left(  \varepsilon\right)  \propto\exp\left(  -\beta
\varepsilon\right)  $ yields:%
\begin{equation}
p\left(  x\right)  =1-\frac{\exp\left(  x\right)  \Phi\left(  x\right)
}{I_{0}\left(  x\right)  }, \label{p.eq}%
\end{equation}
where $x=\beta m$, $I_{0}\left(  x\right)  $ the modified Bessel function of
zero-order and $\Phi\left(  x\right)  $ is given by:%
\begin{equation}
\Phi\left(  x\right)  =\frac{4}{\pi}\sqrt{\frac{2x}{\pi}}\int_{0}^{1}%
\exp\left(  -\frac{2x}{y^{2}}\right)  K\left(  y\right)  \frac{dy}{y^{2}},
\end{equation}
being $K\left(  x\right)  $ the complete Legendre elliptic integral of the
first kind:%
\begin{equation}
K\left(  x\right)  =\int_{0}^{\pi/2}\frac{dt}{\sqrt{1-x^{2}\sin^{2}t}}.
\end{equation}

The energetic dependence of the relative population $p$ of the clustered phase
is illustrated in FIG.\ref{population.eps}. As expected, the system
experiences an abrupt formation of the clustered phase below of the critical
point $u_{c}$. Interestingly, the transformation of the clustered phase into
the gaseous one mainly takes place on the anomalous energy interval
$u_{1}<u\leq u_{c}$, a result that reinforces the hypothesis about the origin
of anomalous behavior in that region should be related to the occurrence of a
dynamical phase coexistence during the out-of-equilibrium regime.%

\begin{figure}
[t]
\begin{center}
\includegraphics[
height=2.5816in,
width=3.5113in
]%
{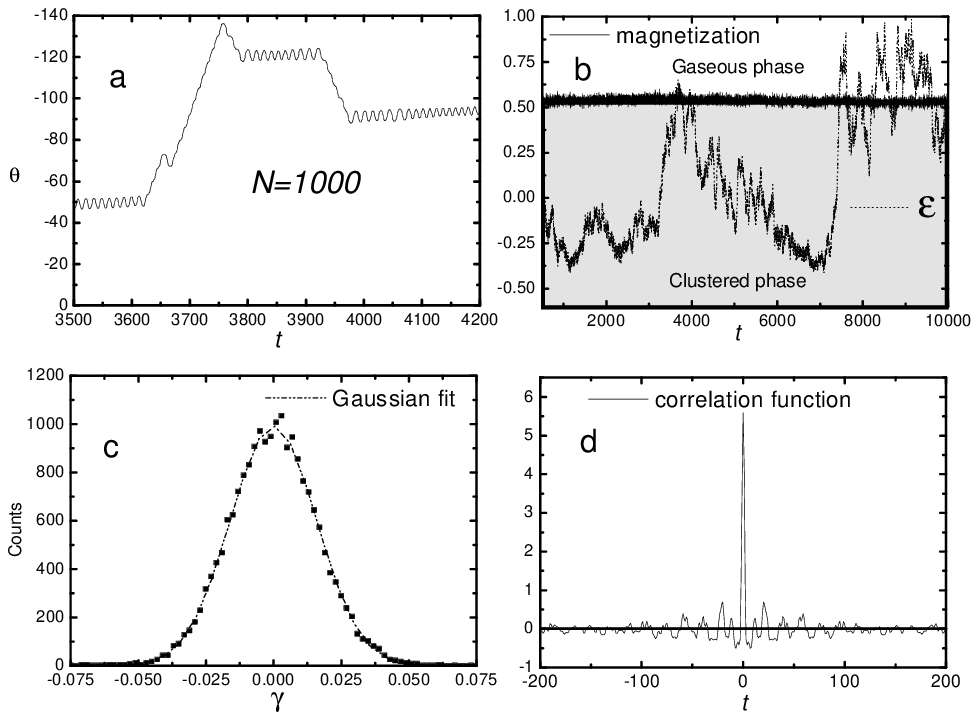}%
\caption{"Particle interchange" between the clustered and the gaseous phases.
Panel a: Transition between vibrational and rotational motions during the
quasi-stationary regime; Panel b: Evolution of the energy $\varepsilon\left(
t\right)  $ of a given particle and the magnetization $m\left(  t\right)  $;
Panels c and d: Frecuency counts and correlation function of the rate of
change of the particle energy $\gamma\left(  t\right)  =d\varepsilon\left(
t\right)  /dt$.}%
\label{dyn.coexist.eps}%
\end{center}
\end{figure}

Panel a) in FIG.\ref{dyn.coexist.eps} shows that a given rotator\ randomly
changes the character of its motion during the quasi-stationary regime, a
phenomenon which could be interpreted here as a "particle interchange" between
the clustered and the gaseous phases. A better understanding of this process
is achieved by analyzing the evolution of the particle energy $\varepsilon
=\varepsilon\left(  \theta,p\right)  =\frac{1}{2}p^{2}-\mathbf{m\cdot
m}\left(  \theta\right)  $ and the magnetization $m=\left\vert \mathbf{m}%
\right\vert $ illustrate in panel b) of this figure. Transitions towards the
gaseous phase take place wherever the dependence $\varepsilon\left(  t\right)
$ overcome the magnetization $m\left(  t\right)  $. Although the dynamical
evolution of magnetization and the particle energy are actually smooth at the
microscopic timescale $\tau_{0}$, these evolution look very rough at a larger
timescale. It is very interesting to remark that the evolution of rate of
change of the particle energy $\gamma\left(  t\right)  =d\varepsilon\left(
t\right)  /dt$ seems to be modeled as a \textit{Gaussian noise} with a
correlation time $\tau_{c}$ comparable to the microscopic timescale $\tau_{0}%
$, as shown in panels c) and d).

\section{Phenomenological approach of collisional relaxation\label{pheno}}

Recently, some authors are carrying out some pioneering studies in order to
obtain some kinetic equations able to explain how this model system evolutes
towards the final equilibrium configurations
\cite{chava,chava2,chava3,bouchet1,bouchet2}. While the early relaxation
dynamics is very-well described by the Vlasov dynamics (\ref{vd}), no one of
such attempts satisfactorily explains the dynamical behavior of the HMF model
in during the collisional regime.

This aim could be achieved by specifying the dynamical mechanism allowing the
energetic interchange among the system particles during the collisional
regime. As already commented in the previous section, the energetic
interchange among the rotators during the collisional regime takes place as
consequence of small oscillations or fluctuations of the magnetization vector
$\mathbf{m}$ around the mean value $\left\langle \mathbf{m}\right\rangle $.
The characteristic timescale of such fluctuations is comparable to the
microscopic timescale $\tau_{0}$, which is the characteristic\ timescale of
the rotator individual evolution. This latter observation and the form of
Eq.(\ref{oscilator}) support the idea that the energy interchange in the HMF
model takes place by means of the \textit{parametric resonance }%
\cite{landau,pet1,pet2,pet3,sos}.

The mechanism of parametric resonance potentially appears whenever exist an
oscillatory dependence of the microscopic parameters of an oscillatory system
\cite{landau}. According with the common understanding, this phenomenon can
manifest in the dynamics (\ref{oscilator}) when a Fourier mode $\omega$ of the
magnetization $\mathbf{m}\left(  t\right)  =\left\langle \mathbf{m}%
\right\rangle +\sum_{\omega}\mathbf{a}_{\omega}\left(  t\right)  \cos\left(
\omega t+\delta_{\omega}\right)  $ belongs to a very small neighborhood of the
frequency $\omega_{i}=2\pi/T_{i}$ of a given rotator: $\left\vert
\omega-\omega_{i}\right\vert \leq\left\vert \mathbf{a}_{\omega}\right\vert
/\omega_{0}\sim1/\sqrt{N}$, being $\omega_{0}\simeq\sqrt{\left\langle
m\right\rangle }$.

At low energies, most of the system rotators exhibit a harmonic oscillatory
motion whose frequencies are very close to $\omega_{0}$. Under these
conditions the parametric resonance allows an affective energy interchange
among the particles leading the system to a fast equilibration, which explains
the concordance of thermodynamical predictions with the microcanonical
numerical experiments at low energies \cite{lat4}. The growing of the energy
per particle leads to the increasing of the non harmonic character of the
rotator motion, and consequently, the spreading of the Fourier modes of
$\mathbf{m}\left(  t\right)  $. Thus, the necessary conditions for the
parametric resonance are only eventually satisfied, allowing in this way to a
given rotator to gain or lost a certain amount of energy, sometimes
significant, in a brief time period, as already illustrated in
FIG.\ref{dyn.coexist.eps}. Such large energetic transfers are rather analogous
to a close encounter.

Recently studies about the origin of the Hamiltonian chaos by using the
methods derived from the Riemannian interpretation of dynamics conclude that
the phenomenon of parametric resonance is the fundamental mechanism to induce
irregular motion in most of the Hamiltonian systems which are binding for most
of regions of the configurational space \cite{pet1,pet2,pet3,sos}. Apparently,
the parametric resonance is the dynamics mechanics explaining the origin of
dynamics instability and the collisional relaxation of the HMF model.

When the dynamical behavior of the system parameters is very complex, the
treatment of such a\ microscopic dynamics could be performed by using the
methods of theory of \textit{random flights} (see in ref.\cite{pet3} the
application of such methods in the framework of the Hamiltonian chaos). This
idea follows from the observation that the small oscillatory behavior of the
magnetization vector in timescale longer than the microscopic time $\tau_{0}$
may be taken into account as a \textit{Gaussian noise}. In fact, the dynamical
evolution of the rotator energy $\varepsilon=\frac{1}{2}p^{2}-\mathbf{m}%
\cdot\mathbf{m}\left(  \theta\right)  $ illustrated in
FIG.\ref{dyn.coexist.eps} suggests that the rotator dynamics is rather
analogous to a \textit{Brownian motion}, that is, a diffusion-like process.

Let us apply the present hypothesis in order to obtain a phenomenological
kinetic equation for describing the collisional evolution of the HMF model.
Hereafter, the Einstein summation convention is assumed. The magnetization
vector $\mathbf{m}$ could be decomposed by considering the its mean value
$\left\langle \mathbf{m}\right\rangle $ plus a \textit{Gaussian noise} term as follows:%

\begin{equation}
\mathbf{m}\simeq\left\langle \mathbf{m}\right\rangle +\mathbf{n}_{a}\eta_{a},
\end{equation}
where the unitary vectors $\mathbf{n}_{a}$ consider a different behavior of
the noises $\eta_{a}$ in the parallel and transverse direction of the mean
value $\left\langle \mathbf{m}\right\rangle $. Let us also assume that the
noise only depends on the global system observables (system size $N$, energy
$u$,\ average magnetization vector $\left\langle \mathbf{m}\right\rangle $,
etc.). Since a purely diffusive influence does not lead the one-body
distribution function towards the Gaussian profiles of velocities resulting at
the final equilibrium, we shall take into consideration the inclusion of a
\textit{dynamical friction} in this context in order to accomplish this
feature\textit{ }\cite{chandra}.

Thus, we conjecture that the microscopic dynamics of a given rotator could be
conveniently approximated by the following Langevin equation with a
\textit{multiplicative noise}:%

\begin{equation}
\dot{\theta}=p,~\dot{p}=-\lambda\left(  \theta\right)  p+\mathbf{m}\left(
\theta\right)  \mathbf{\times}\left\langle \mathbf{m}\right\rangle
+e_{a}\left(  \theta\right)  \eta_{a}. \label{ansatz}%
\end{equation}
where the dynamical friction $\lambda\left(  \theta\right)  $ should
necessarily be $\theta$-dependent (see in Eq.(\ref{einstein}) below). The
functions $e_{a}\left(  \theta\right)  $ are derived from the unitary vectors
$\mathbf{n}_{a}$ as $e_{a}\left(  \theta\right)  =\mathbf{m}\left(
\theta\right)  \times\mathbf{n}_{a}$. The Gaussian noises $\eta_{a}$ satisfy
the following relations:%

\begin{equation}
\left\langle \eta_{a}\left(  t\right)  \right\rangle =0,~\left\langle \eta
_{a}\left(  t\right)  \eta_{b}\left(  t^{\prime}\right)  \right\rangle
=2\Omega_{ab}\delta\left(  t-t^{\prime}\right)  . \label{corr.matrix}%
\end{equation}
where $\Omega_{ab}$ is the correlation matrix. The Langevin equation
(\ref{ansatz}) leads to the following Fokker-Planck equation:%

\begin{equation}
\frac{\partial f}{\partial t}+p\frac{\partial f}{\partial\theta}%
+\mathbf{m}\left(  \theta\right)  \mathbf{\times m}\left[  f\right]
\frac{\mathbf{\partial}f}{\partial p}=D\left[  \theta\right]  \frac{\partial
}{\partial p}\left[  \frac{\partial f}{\partial p}+\eta pf\right]  ,
\label{fokker}%
\end{equation}
where the mean value of the magnetization vector $\left\langle \mathbf{m}%
\right\rangle $ was taken as follows:%

\begin{equation}
\left\langle \mathbf{m}\right\rangle =\mathbf{m}\left[  f\right]  =\int
d\theta dp~\mathbf{m}\left(  \theta\right)  f, \label{mean}%
\end{equation}
and the diffusion coefficient $D\left[  \theta\right]  $ by:%

\begin{equation}
D\left[  \theta\right]  =\Omega_{ab}e_{a}\left(  \theta\right)  e_{b}\left(
\theta\right)  . \label{dif.function}%
\end{equation}
We also assume that $D\left[  \theta\right]  $ is related to the dynamical
friction by means of the Einstein relation:%

\begin{equation}
\lambda\left(  \theta\right)  =\eta D\left(  \theta\right)  , \label{einstein}%
\end{equation}
being $\eta$ the nonequilibrium analogue of the inverse temperature,
$\eta^{-1}\equiv\int p^{2}f\left(  \theta,p;t\right)  dpd\theta$. The
consistence of Fokker-Planck equation (\ref{fokker}) demands that the
correlation matrix $\Omega_{ab}$ should be considered as certain functionals
of the distribution function $f$. Since $\Omega_{ab}$ decreases with the
increasing of the system size, the diffusive term of Fokker-Planck equation is
only effective at large temporal scales. Thus, the assumption (\ref{mean})
takes into account that the dynamical evolution of the distribution function
at the timescale $\tau_{0}$ is described by the collisionless dynamics
(\ref{vd}), while diffusive term considers the relaxation towards the
Boltzmann-Gibbs distribution function at the collisional timescale $\tau
_{cr}\propto D^{-1}$.

The definition of the diffusion coefficient $D\left(  \theta\right)  $ can be
conveniently rewritten by considering the \textit{tridimensional} vector
$\mathbf{m=}\left(  \cos\theta,\sin\theta,0\right)  $. Thus, the scalar term
$e_{a}e_{b}$ could be rephrased as:%

\begin{equation}
e_{a}e_{b}\equiv\left(  \mathbf{m}\times\mathbf{n}_{a}\right)  \cdot\left(
\mathbf{m}\times\mathbf{n}_{b}\right)  =\delta_{ab}-\left(  \mathbf{m}%
\cdot\mathbf{n}_{a}\right)  \left(  \mathbf{m}\cdot\mathbf{n}_{a}\right)  .
\end{equation}
Let us now reconsider again the bidimensional form of the vector
$\mathbf{m}\left(  \theta\right)  =\left(  \cos\theta,\sin\theta\right)  $. By
denoting the \textit{i}-th component of $\mathbf{m}\left(  \theta\right)  $ by
$m_{i}\left(  \theta\right)  $, the diffusion coefficient can be expressed as:%

\begin{equation}
D=m_{i}\left(  \theta\right)  \mathcal{K}_{ij}m_{j}\left(  \theta\right)  ,
\end{equation}
where $\mathcal{K}_{ij}=\Omega\delta_{ij}-\Omega_{ab}\left(  n_{a}\right)
_{i}\left(  n_{b}\right)  _{j}$, and $\Omega=Sp\left[  \Omega_{ab}\right]  $.
Since the only preferential direction of this system is the mean value of the
magnetization $\left\langle \mathbf{m}\right\rangle $, the symmetric matrix
$\mathcal{K}_{ij}$ should take the form:%

\begin{equation}
\mathcal{K}_{ij}=a\left[  f\right]  \delta_{ij}+b\left[  f\right]
\left\langle m_{i}\right\rangle \left\langle m_{j}\right\rangle ,
\end{equation}
allowing to rewrite the diffusion factor as follows:%
\begin{equation}
D\left(  \theta\right)  =a\left[  f\right]  +b\left[  f\right]  \left\{
\mathbf{m}\left[  f\right]  \cdot\mathbf{m}\left(  \theta\right)  \right\}
^{2}.
\end{equation}
The closure of the Fokker-Planck approximation relies on the determination of
the functional $a\left[  f\right]  $ and $b\left[  f\right]  $, which is a
task beyond of our phenomenological reasonings. It is very easy to verify that
Eq.(\ref{fokker}) ensures the particles, momentum and the energy conservation.

Chavanis \textit{et al} introduced in refs.\cite{chava,chava2,chava3} a
similar treatment by modifying of the HMF model microscopic dynamics
(\ref{motion equations}) with the inclusion of\ frictional and stochastic
terms, whose resulting Fokker-Planck equation\ basically differs from the
present proposal by the $\theta$-dependence of the diffusion coefficient
$D\left(  \theta\right)  $. The origin of this difference is found in the fact
that these authors actually introduce a variant of HMF model (Brownian Mean
Field model) which accounts for the external influence of a thermostat, while
the frictional and stochastic terms considered in the present work try to
capture the effective influence of the magnetization fluctuations on the exact
individual dynamics of a given rotator in a timescale larger than the
microscopic time $\tau_{0}$. The crucial question now is how to justify and
precise within the framework of the parametric resonance the \textit{ad hoc}
approximation (\ref{ansatz}), mainly, the inclusion of the effective dynamical friction.

\section{Final remarks}

As already appreciated in many studies, the HMF model dynamics undergoes the
incidence of two relaxation process with different characteristic timescales:
(1) the violent relaxation timescale $\tau_{vr}$ leading to the
quasi-stationary evolution with a superdiffusion regime, and (2) the
collisional relaxation timescale $\tau_{cr}$ where the Boltzmann-Gibbs
equilibrium is achieved as final state. It is well-established theoretically
and numerically that the collisionless relaxation timescale is the same order
of the microscopic timescale, $\tau_{vr}\sim\tau_{0}$, while the collisional
relaxation timescale seems to grow with the system size in some power of $N$,
$\tau_{cr}\sim\tau_{0}N^{\alpha}$. Early estimations predict a simple linear
dependence \cite{vrt}, although there are also several numerical experiments
evidencing a nontrivial exponent $\alpha\simeq1.7$
\cite{zanette,yamaguchi,ybd}.

Generally speaking, the determination of the correct \textit{N}-dependence of
the collisional relaxation time is rather difficult to carry out by means of
numerical computations of the microscopic dynamics since such experiments are
plagued of prejudicial size effects and poor equilibration of trajectory
averages whose elimination seems to demand the consideration of more larger
number of particles than the actual computational possibilities $N\sim
10^{4}-10^{5}$ \cite{zanette}. For example, besides of the nontrivial exponent
$\alpha=1.7$ of the power-law growing of $\tau_{cr}$ reported by Yamaguchi
\textit{et al} in refs.\cite{yamaguchi,ybd}, they also indicated the following
\textit{N}-dependence of the\ quantity $q$:%
\begin{equation}
q=\left.  \frac{d\left\langle m\left(  t\right)  \right\rangle }{d\left(  \ln
t\right)  }\right\vert _{t=\tau_{cr}}\propto\sqrt{N}. \label{r1}%
\end{equation}
If the above estimation of $\tau_{cr}$ is the correct relaxation timescale
associated to the collisional quasi-stationary evolution (qse) of the one-body
distribution function $f\left(  \theta,p;t\right)  $, the temporal dependence
for large $N$ should be given by:
\begin{equation}
f\left(  \theta,p;t\right)  \simeq f_{qse}\left(  \theta,p;t/\tau_{cr}\right)
+O\left(  1/N\right)  .
\end{equation}
The above assumption implies that the leading behavior the temporal dependence
of the average magnetization should be given by $\left\langle m\left(
t\right)  \right\rangle \simeq A\left(  t/\tau_{cr}\right)  $, where $A\left(
s\right)  =\left\vert \int\mathbf{m}\left(  \theta\right)  f_{0}\left(
\theta,p;s\right)  d\theta dp\right\vert $ is a size-independent function.
Thus, the leading behavior of the quantity:%
\begin{equation}
q=\left.  \frac{d\left\langle m\left(  t\right)  \right\rangle }{d\ln
t}\right\vert _{t=\tau_{cr}}\equiv\frac{dA\left(  s=1\right)  }{ds},
\end{equation}
\textit{should be }$N$-\textit{independ }when $N$ is large enough. Since the
numerical result (\ref{r1}) disagrees with this reasoning, the nontrivial
exponent $\alpha\simeq1.7$ reported in the Yamaguchi \textit{et al} study
might not correspond to the true scaling behavior for $N$ very large, but only
a finite size behavior which should disappear for $N$ large enough.

The size dependence of the collisional relaxation timescale is a question with
a primordial importance in order to understand the dynamical features of a
long-range interacting system as the HMF model when thermodynamic limit is
invoked. As already pointed out by some authors, the existence of relaxation
timescale $\tau_{cr}$ diverging with the imposition of the thermodynamic
limit, $\lim_{N\rightarrow\infty}\tau_{cr}=\infty$, leads automatically to the
non commutative character of the the thermodynamic limit $\lim_{N\rightarrow
\infty}$ with the infinite time limit $\lim_{t\rightarrow\infty}$ necessary to
the equilibration of temporal averages $\left\langle A_{N}\right\rangle
_{t}=\int_{0}^{t}A_{N}\left(  \tau\right)  d\tau/t$:
\begin{equation}
\lim_{N\rightarrow\infty}\lim_{t\rightarrow\infty}\frac{\left\langle
A_{N}\right\rangle _{t}}{N}\not =\lim_{t\rightarrow\infty}\lim_{N\rightarrow
\infty}\frac{\left\langle A_{N}\right\rangle _{t}}{N}, \label{non.comm}%
\end{equation}
Thus, the imposition of the thermodynamic limit before the infinite time limit
makes endless the duration of collisionless dynamics, and therefore, the
Boltzmann-Gibbs equilibrium arises as an admissible but particular
quasi-stationary state of the violent relaxation. Such a dynamical picture is
not only a feature of the HMF model dynamics. On the contrary, the large the
collisional relaxation times estimates of many astrophysical objects like the
elliptical galaxies ($\tau_{cr}\sim0.1\tau_{0}N/\ln N$ \ where microscopic
timescale $\tau_{0}\sim1/\sqrt{G\rho}$) induces also to suppose the
collisionless character of the dynamical evolution of such real long-range
interacting systems, which explains the rich variety of structures observed in
this context \cite{bin}.

Let us use the phenomenological approach of collisional regime in terms of the
Fokker-Planck equation (\ref{fokker}) in order to analyze the \textit{N}%
-dependence of the collisional relaxation timescale $\tau_{cr}$. Since the
amplitude of the Gaussian noises $\eta_{a}$ describing the magnetization
fluctuations during the quasi-stationary evolution decreases as $1/\sqrt{N}$
as well as the underlying correlation times are of order of microscopic
timescale $\tau_{0}$ (see in panel c) of FIG.\ref{dyn.coexist.eps}), the
correlation matrix $\Omega_{ab}$ (\ref{corr.matrix}) and the diffusion
coefficient $D\left(  \theta\right)  $ (\ref{dif.function}) decrease as $1/N$.
Thus, the phenomenological picture described in this work suggests that the
relevant timescale of the collisional evolution of the HMF model should obey a
linear growing $\tau_{cr}=\tau_{0}N$.

As already discussed in our previous work \cite{vel.HMF.1}, such a linear
\textit{N}-dependence of $\tau_{cr}$ seems to be intimately related to the
\textit{N}-dependence of the additive constant of entropy per particle
$s_{0}=\frac{1}{2}\ln\left(  2\pi e^{2}Ig/N\right)  $ , since the imposition
of a scaling dependence of the coupling constant $g$ as $g\left(  N\right)
=\gamma N$ instead of the usual Kac prescription \cite{kac} $g_{kac}\left(
N\right)  =\gamma/N$ leads to a \textit{simultaneous regularization} of the
divergence of the collisional relaxation timescale $\tau_{cr}=\tau_{0}%
N=\sqrt{IN/g}\equiv\sqrt{I/\gamma}$ and the additive constant $s_{0}=\frac
{1}{2}\ln\left(  2\pi e^{2}Ig/N\right)  \equiv\frac{1}{2}\ln\left(  2\pi
e^{2}I\gamma\right)  $ in the thermodynamic limit \cite{vel.HMF.1}, avoiding
in this way the incidence of an undesirable dynamical anomalies like the one
described in Eq.(\ref{non.comm}) and the divergence of the thermodynamic potentials.

Although the phenomenological approach of collisional relaxation described in
this work supports the validity of the linear growing $\tau_{cr}=\tau_{0}N$,
the problem about the exact \textit{N}-dependence of collisional relaxation
timescale could be definitively solved by means of the development of
appropriate kinetic equations starting from the consideration of first
principles \cite{chava3}. We shall address in our forthcoming paper an intense
investigation of the HMF model dynamics in terms of kinetic equations, where
the implementation of suitable kinetic equations starting from the well-known
\textit{BBGKY hierarchy} will receive a primordial attention.


\begin{thebibliography}{99}                                                                                               %


\bibitem {kk}T. Konishi and K. Kaneko, J. Phys. A \textbf{25} (1992) 6283.

\bibitem {pichon}C. Pichon, PhD thesis, Cambridge (1994).

\bibitem {inaga}S. Inagaki, Prog. Theor. Phys. \textbf{96} (1996) 1307.

\bibitem {ant}M. Antoni and S. Ruffo, Phys. Rev. E \textbf{52} (1995) 2361.

\bibitem {lat1}V. Latora, A. Rapisarda and S. Ruffo, Phys. Rev. Lett.
\textbf{83} (1999) 2104; Physica A \textbf{280} (2000) 81.

\bibitem {lat2}V. Latora, A. Rapisarda and S. Ruffo, Physica D \textbf{131}
(1999) 38; e-print (1998) [chao-dyn/9803019].

\bibitem {lat3}V. Latora and A. Rapisarda, Prog. Theor. Phys. Suppl.
\textbf{139} (2000) 204.

\bibitem {lat4}V. Latora, A. Rapisarda and S. Ruffo, Nucl. Phys. A
\textbf{681} (2001) 331c.

\bibitem {vrt}V. Latora, A. Rapisarda and C. Tsallis, Phys. Rev. E \textbf{64}
(2001) 056134; Physica A \textbf{305} (2002) 129.

\bibitem {zanette}D. H. Zanette and M. A. Montemurro, Phys. Rev. E \textbf{67}
(2002) 031105.

\bibitem {dauxois}T. Dauxois, S. Ruffo, E. Arimondo and M. Wilkens (Eds),
\textit{Dynamics and thermodynamics of systems with long range interactions},
Lecture Notes in Physics (Springer, 2002) and ref therein.

\bibitem {yamaguchi}Y.Y. Yamaguchi, Phys. Rev. E \textbf{68} (2003) 066210.

\bibitem {ybd}Y.Y. Yamaguchi, J. Barre, F. Bouchet, T. Dauxois, and S. Ruffo,
Physica A \textbf{337} (2004) 36.

\bibitem {chava}P.H. Chavanis, J. Vatteville and Bouchet, Eur. Phys. J. B
\textbf{46}(2005) 61; e-print (2004) [cond-mat/0408117].

\bibitem {chava2}P.H. Chavanis, Phys. Rev. E \textbf{68} (2003) 036108;
e-print(2002) [cond-mat/0209096].

\bibitem {chava3}P.H. Chavanis, Physica A \textbf{361} (2006) 81; e-print
(2004) [cond-mat/0409641v3].

\bibitem {bouchet1}F. Bouchet, e-print(2003) [cond-mat/0305171].

\bibitem {bouchet2}F. Bouchet and T. Dauxois, e-print (2004) [cond-mat/0407703].

\bibitem {vel.HMF.1}L. Velazquez and F. Guzman, \textit{Remarks about the
thermostatisical description of the HMF model. Part I: Equilibrium
Thermodynamics}

\bibitem {bin}J. Binney and S. Tremaine, \textit{Galactic Dynamics} (Princeton
Series in Astrophysics, Princeton, NJ, 1987).

\bibitem {landau}L.D. Landau and E.M. Lifshitz, \textit{Course Of Theoretical
Physics:} \textit{Mechanics} (Pergamon Press, Oxford, 1960).

\bibitem {pet1}M. Pettini, Phys. Rev. E \textbf{47} (1993) 828; M.Cerruti-Sola
and M. Pettini, Phys. Rev. E \textbf{53} (1996).

\bibitem {pet2}P. Cipriani and M. Pettini, Astrophys. Space Sci. \textbf{283}
(2003) 347; e-print (2001) [astro-ph/0102143].

\bibitem {pet3}L. Casetti, C. Clementi and M. Pettini, Phys. Rev. E
\textbf{54} (1996) 5969.

\bibitem {sos}R. Sospedra-Alfonso, L.Velazquez and J.Rubayo-Soneira, Chem.
Phys. Lett. \textbf{375} (2003) 261.

\bibitem {chandra}S. Chandrasekhar, Astrophys. J. \textbf{98} (1943) 54;
\textit{Principles of Stellar Dynamics }(Dover Publications Inc., New York, 1960).

\bibitem {kac}M. Kac, G.E. Uhlenbeck and P.C. Hemmer, J. Math. Phys.
\textbf{4} (1963) 216.
\end{thebibliography}
\end{document}